# Systimator: A Design Space Exploration Methodology for Systolic Array based CNNs Acceleration on the FPGA-based Edge Nodes

Hazoor Ahmad[1], Muhammad Tanvir[1], Muhammad Abdullah Hanif[2], Muhammad Usama Javed[1], Rehan Hafiz[1], Muhammad Shafique[2]

[1]Department of Electrical Engineering, Information Technology University, Lahore, Pakistan
[2]Department of Computer Engineering, Vienna University of Technology, 1040 Wien, Austria

*Abstract—* The evolution of IoT based smart applications demand porting of artificial intelligence algorithms to the edge computing devices. CNNs form a large part of these AI algorithms. Systolic array based CNN acceleration is being widely advocated due its ability to allow scalable architectures. However, CNNs are inherently memory and compute intensive algorithms, and hence pose significant challenges to be implemented on the resource-constrained edge computing devices. Memory-constrained low-cost FPGA based devices form a substantial fraction of these edge computing devices. Thus, when porting to such edge-computing devices, the designer is left unguided as to how to select a suitable systolic array configuration that could fit in the available hardware resources. In this paper we propose Systimator, a design space exploration based methodology that provides a set of design points that can be mapped within the memory bounds of the target FPGA device. The methodology is based upon an analytical model that is formulated to estimate the required resources for systolic arrays, assuming multiple data reuse patterns. The methodology further provides the performance estimates for each of the candidate design points. We show that Systimator provides an in-depth analysis of resource-requirement of systolic array based CNNs. We provide our resource estimation results for porting of convolutional layers of TINY YOLO, a CNN based object detector, on a Xilinx ARTIX 7 FPGA.

## I. INTRODUCTION AND RELATED WORK

Owing to their recent success in a variety of complex computer vision and image classification challenges, Convolutional Neural Networks (CNNs) have found their widespread application in the areas of real time video surveillance, autonomous driving, natural language processing, robotics and more [1][2]. CNNs involve multiple convolutional layers, whereby an input feature map (IFM) is convolved with multiple filters (defined by their weights) to generate an output that is further pooled and passed through an activation function to generate the corresponding output feature map (OFM). CNNs are inherently memory and compute intensive algorithms and hence their hardware acceleration has received a lot of interest. Recent advances in their hardware acceleration has mostly focused on maximizing the design reuse [3] for weights [4][5] and/or the input/output feature maps [6][7]. Recently, Systolic Array (SA) based accelerators are being widely employed because of their scalable and flexible structure, and convolution friendly nature [10-12]. Zhang et. al. [10] provided one of the seminal study in this area by proposing an analytical design scheme to quantitatively estimate the computing throughput and required memory bandwidth of a CNN for a particular FPGA. They made use of loop tiling to maximize the design reuse for their proposed Systolic Array based architecture. However, since all the processing elements (PE) of their architecture were directly connected to the on-chip memory, their design did not conform to a strict systolic array architecture that exploits the local interconnects [12]. To overcome this issue, a 1-D systolic array design for AlexNet was proposed by [11] which provided a higher throughput. Unfortunately, they assumed all the IFMs to be stored in the on-chip memory, and hence their architecture is not suitable for memory constrained FPGAs that are typically employed in edge devices. Wei et. al. [12] provided exhaustive search for design space exploration (DSE) of a fully systolic array based architecture for FPGAs. However, unlike Caffeine [10], their architecture only considers feature map reuse and do not consider the filter weight reuse. Furthermore, both [10] and [12] focus on searching a high throughput design for high performance FPGAs that can typically host a decent enough systolic array. When considering low-cost FPGAs as the computing unit for edge computing, the on-chip resources are limited. As an example consider an Artix7 FPGA with 86K logic slices, 220 DSP units, and 4.9 Mb of block RAM as compared to Kintex Ultrascale (331.68K logic slices, 2760 DSP units, and 38.0 Mb of block RAM) which was targeted by[10]. Thus, when moving to such resource-constrained FPGAs, the problem of finding a suitable systolic array size that could fit in within the resource bounds is a challenging task.

**Limitations of state of the art:** State of the art design space exploration schemes for enabling CNNs on FPGAs either employ systolic arrays that do not completely employ the local interconnections among the PEs [10], or do not consider multiple data traversal orders [12]. Furthermore, their goal has been to provide a high throughput design intended for larger FPGAs. For low-cost memory-constrained edge-devices, FPGAs may have much smaller number of DSP slices and on-chip memory. Thus, when porting to such edge-computing devices, the designer is left unguided as to how to select a suitable systolic array configuration that could fit in the available hardware resources.

**Required:** Thus, there is a requirement for a combined resource and performance estimation based design space exploration methodology that could evaluate multiple design points that relate to a variety of systolic array sizes for the device under consideration.

**Our Contribution:** In this paper we present Systimator, a design space exploration methodology that is based upon a tile based systolic array architecture which supports multiple data traversal orders. Provided the CNN's layer wise network information and the hardware constraints, Systimator explores a wide design space, based upon the proposed layer wise analytical models for resource estimation. It provides a list of valid design points, which correspond to various systolic array dimensions and tile sizes, which could fit within the resource constraints of the FPGA being considered. Furthermore, it provides the performance ranking of the extracted design points using a proposed model for performance estimation. The methodology is explored for porting of convolutional layers of Tiny YOLO object detection [13] network on Artix7 FPGA. The scripts for the Systimator methodology shall be provided as open source at [14] to facilitate further research and development.

## II. SYSTIMATOR : DESIGN SPACE EXPLORATION BASED RESOURCE ESTIMATION FOR CNN IMPLEMENTATION ON SYSTOLIC ARRAYS

First we present a systolic array (SA) based architecture design for modelling our Systimator design space exploration. The generic nature of this architecture allows our methodology to be applied on a variety of existing systolic array architectures such as in [12]. Our proposed architecture is illustrated in Figure 1 and described below.

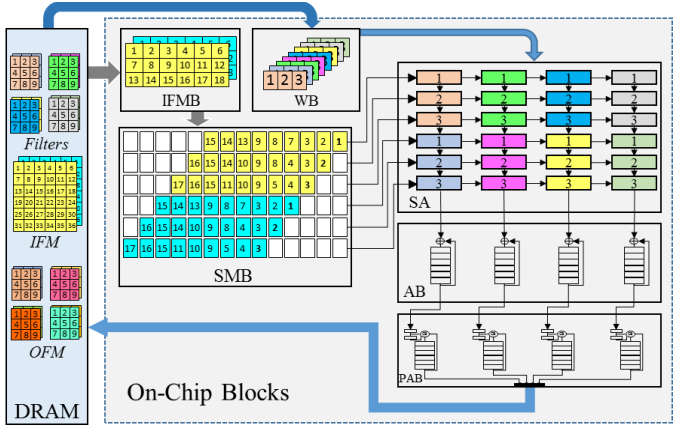

Figure 1: Systolic Array based architecture for Systimator. It comprises the IFMB (Input Feature Map Buffer), WB (Weight Buffer), SMB (Scratchpad Memory Buffer), SA (Systolic Array), AB (Accumulation Block), PAB (Pooling and Activation Block).

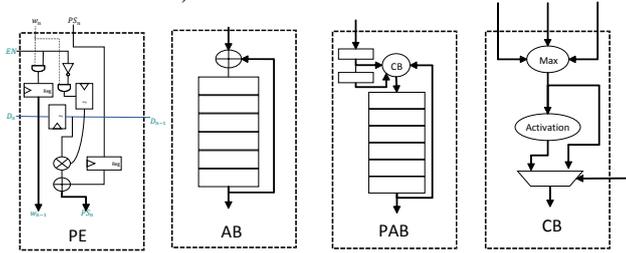

Figure 2: Architectural details of various on-chip blocks.

### A. Accelerator Architecture for Systimator

Since, IFMs and weights of a CNN require large storage, they are stored in an off-chip DRAM. Figure 1 illustrates a case where the IFM comprise two channels of a 6x6 feature map. A tile of IFM and the corresponding filter weights are brought in the on-chip IMF buffer (IFMB) and weight buffer (WB), respectively. For the illustration in Figure 1 the tile is composed of three rows of IFM. There is one row of scratchpad memory corresponding to every row of systolic array. The tile data from the IFMB is sequenced appropriately in the scratchpad memory (SMB), as directed by the data traversal order, for onward transfer to the systolic array (SA). We consider two different types of data traversals:

1. *Feature map reuse*: Next tile data is not fetched unless the current tile data has been completely consumed by all the filters of a specific CNN layer being processed.

2. *Filter reuse:* Systolic Array filters are not updated unless all the tiles of an IFM have been processed by current set of SA filters.

#### 1) Systolic Array (SA)

On every clock cycle, IFM data travels rightward and partial sums move downwards. Each column of the SA corresponds to a row of filter being operated on the input IFM for multiple channels. In Figure 1, two channels of IFM are being processed by two channels of first filter. Each PE multiplies both its inputs, adds it with partial sum obtained from its neighboring top PE, and passes the latest accumulation of product to the PE below. The systolic array is operated multiple times to process all the rows of the filter. Once the partial sums through each column of systolic array are available at the output of its last row, they are fed into the accumulation block as shown in Figure 1.

#### 2) Accumulation Block (AB)

This block is composed of a FIFO (to store partial sums) and an adder to add the incoming partial sum with its respective counterpart already stored in that block. This time synchronized addition continues until we have the completely convolved output present in each block. Each block's contents are frozen when there is no relevant data incoming from the systolic array.

#### 3) Pooling and Activation Block (PAB)

Contents of each AB, when ready, are pushed into its respective pooling and activation block. Each such block is composed of a comparator block, a memory element (FIFO) and an activation unit. Successive inputs (two, three etc. based on size of pooling kernel) from accumulation block are compared with their corresponding entry already stored in the residual FIFO of this block and the largest among these values is the output of comparator block. Once the output is completely pooled it is then fed to the activation unit which applies one of three activations on this data i.e. ReLU, Leaky ReLU, ELU, and resultant output is again stored in the same FIFO. Based on the data traversal order the depth of all included FIFOs may vary. Finally the OFM is transferred back to the DRAM.

### B. Systimator based Design Space Exploration

In Table I, we define the set of parameters being used by our Systimator model to describe the CNN network, the hardware constraints and the design space points being explored.

Table I: Systimator Parameters

| CNN Network Parameters for an L layer network ($l = 1, 2, ... L$) | | | |
|---|---|---|---|
| Total # of layers of a CNN | $L$ | # of Rows of $l_{th}$ layer's IFM | $r(l)$ |
| # of filters in $l_{th}$ layer | $n_f(l)$ | # of Cols. of $l_{th}$ layer's IFM | $c(l)$ |
| # of Rows of $l_{th}$ layer filters | $r_f(l)$ | # of ch. of $l_{th}$ layer's IFM | $ch(l)$ |
| # of Cols. of $l_{th}$ layer filters | $c_f(l)$ | Stride in $l_{th}$ layer pooling | $s(l)$ |
| **FPGA/Hardware Design Constraints** | | | |
| # DSP Units | $N_{DSP}$ | Block RAM | $M_{BRAM}$ |
| **Design Parameters for $i_{th}$ design point ($i = 1, 2, ... I$)** | | | |
| # of Rows of SA | $r_{sa}(i)$ | Data traversal (Feature Map) | $\rho(i) = 1$ |
| # of Columns of SA | $c_{sa}(i)$ | Data traversal (Filter Map) | $\rho(i) = 0$ |
| # ch. being processed in SA | $ch_{sa}(i)$ | # of rows for $l_{th}$ layer tile | $r_t(i, l)$ |
| Maximum Tile Size | $M_{tile}(i)$ | # of cols. for $l_{th}$ layer tile | $c_t(i, l)$ |

Let $P$, $Q$ and $R$ be the number of possible set of configurations being explored for selecting the tile size ($r_{sa}$), number of rows of systolic array ($c_{sa}$) and the number of channels ($ch_{sa}$) being processed in parallel by the SA, respectively. Since, a valid design point can be a combination of any of these individual configurations, $I = P \times Q \times R$ is the number of design points being evaluated. A design point $i$ is, thus, uniquely defined by the: systolic array size ($r_{sa}(i) \times c_{sa}(i)$), number of channels being processed in parallel ($ch_{sa}(i)$), the tile size ($r_t(i, l) \times c_t(i, l)$) and the data traversal order $\rho(i)$ being followed. Design space exploration is performed in two steps: Provided the CNN network's parameters, Systimator explores $I$ number of design points and provides a layer wise estimate of the required on-chip memory resources. Only the design points that confirm to the FPGA design constraints ($N_{DSP}, M_{BRAM}$), are considered for the second step. In the second step, the valid design points are subjected to performance evaluation. Each of these two steps are provided below:

*1) Systimator: Resource Estimation*

First, we generate $P$ configurations for the tile size. We start by deciding upon a factor, F, which dictates the maximum tile size such that the maximum number of rows of the IFM tile is bounded by $(r(1)/(F))$. This is justified, since the first layer IFM has the largest number of rows and not all the IFM can be loaded into the on-chip buffer. Candidate tile configurations can thus be generated by varying $r_t(p,l) = \min(([r(1)/(p*F)]), r(l))$, and keeping $c_t(p,l) = c(l)$, for $p = 1,2,3,..P$. Q and R possible values of $c_{sa}$ and $ch_{sa}$ are generated as follows:

$$c_{sa}(q) = 2*q, (\forall\ q = 1,2,..Q) \quad (1)$$

$$ch_{sa}(r) = 2*r, (\forall\ r = 1,2,..R) \quad (2)$$

Thus, we assume a minimum number of 2 columns and 2 channels for our design space exploration. The number of rows of the SA are given by $r_{sa}(r) = ch_{sa}(r) \max_l(r_f(l))$. Any combinations of $r_t(p,l), c_{sa}(q)$ and $ch_{sa}(r)$, thus completely defines an $i^{th}$ design point. Next we compute the layer wise memory requirement for each of the on-chip memory blocks, for each $i^{th}$ design point. Hence, the memory required for IFMB is given by

$$M_{FM}(i,l) = r_t(i,l)c_t(i,l)ch_{sa}(i,l) \quad (3)$$

Here, $r_t(i,l), c_t(i,l)$ and $ch_{sa}(i,l)$ represent the corresponding $r_t, c_t$ and $ch_{sa}$ of the $i_{th}$ design point. The memory required for storing the partial sums, in AB, for $l_{th}$ layer of $i_{th}$ design point, depends upon the data traversal order and is given by:

$$M_{PS}(i,l,\rho) = [(1-\rho)c_{sa}(i) + (\rho)n_f(l)][d_H(i,l)\,d_V(i,l)] \quad (4)$$

Here, $d_H(i,l)\,d_V(i,l)$ is the total number of locations that filter has to be slided in 2-D in order to perform the convolution of a single channel of the IFM. Thus $d_H(i,l) = (r(l) - r_f(l) + 1)$ and $d_V(i,l) = (c(l) - c_f(l) + 1)$. If $s(l)$ is the stride of pooling layer $l$ the memory required for PAB is given by:

$$M_{pool}(i,l,\rho) = \frac{M_{PS}(i,l,\rho)}{s^2(l)} \quad (5)$$

Let, $M_{W_{SA}}(i,l)$ be the minimum amount of memory required to store at-least one set of weights of the systolic array of the $i_{th}$ design point, then the total memory required for the $i_{th}$ design point is given by:

$$M_T(i,l,\rho) = M_{FM}(i,l) + M_{PS}(i,l,\rho) + M_{pool}(i,l,\rho) + M_{W_{SA}}(i,l,\rho) \quad (6)$$

If $M_{BRAM}$ is the effective on-chip memory then, the free memory available after the systolic array implementation is given by:

$$M_\Delta(i,l,\rho) = M_{BRAM} - M_T(i,l,\rho) \quad (7)$$

If $M_\Delta(i,l,\rho)$ is negative, this means design is not feasible. If it is positive, it may be employed to cache extra weight or tile data. Thus, if $\mu(i,l,\rho)$ is the minimum memory for all the $l$ layers of the $i_{th}$ design point, and $\Phi(i,l)$ is the function of feature map size, number of filters, and data traversal order then a valid design configuration is defined by

$$\mu(i,l,\rho) = \min_l M_\Delta(i,l,\rho) \quad (8)$$

$$\Phi(i,l,\rho) = f(ch_{sa}(i), c_{sa}(i), \rho, r_t(i,l), c_t(i,l)) \quad (9)$$

$$\Re = \{\forall\ \Phi(i,l,\rho): \mu(i,l,\rho) > 0\ \&\ n_{dsp} \le N_{dsp}\} \quad (10)$$

where $n_{dsp} = r_{sa}(i)c_{sa}(i)$ is the number of DSP units of design point, and $N_{dsp}$ the number of DSP units available.

*2) Model for Performance Estimation*

Valid design points from the previous step are evaluated for total cycles (compute + memory) that they require to process a complete IFM. The clock cycles required to compute the final result is the sum of clock cycles consumed for IFM data transfer, weight fetching, scratchpad memory filling, systolic array processing and OFM write-back. Since, the purpose is to relativity rank the performance of various design points we simplify our model by making certain assumptions. Firstly, we consider an average throughput of $W$ words/clock cycle as the off-chip memory bandwidth. We do not consider any other kind of DRAM overhead. The IFM tiles that we fetch are non-overlapping. However, in practice, data may be shared among adjacent tiles. Furthermore, our timing assumes sequential memory transfer and computations. In actual, memory and compute operations can be conveniently parallelized. And lastly, we consider an input batch size of 1.

Let $\alpha, \beta$, and $\gamma$ be the tiling factors for filters, IFM rows, and IFM channels respectively, and defined as: $\alpha(i,l) = [n_f(l)/c_{sa}(i)]$, $\beta(i,l) = [r(l)/r_t(i,l)]$ and $\gamma(i,l) = [ch(l)/ch_{sa}(i)]$. And let $\Omega$ be their product, $\Omega(i,l) = \alpha(i,l)\beta(i,l)\gamma(i,l)$. The number of cycles required to bring one tile of IFM from DRAM to IFMB is then given by:

$$T_{FM}(i,l) = \frac{1}{W}(\alpha(i,l)\rho + 1 - \rho)\beta(i,l)\gamma(i,l)M_{FM}(i,l) \quad (11)$$

The number of cycles required to bring corresponding filter from DRAM to $WB$ is given by

$$T_w(i,l) = \frac{1}{W}(\alpha(i,l)(1-\rho) + \rho)\beta(i,l)\gamma(i,l)M_{W_{SA}}(i,l) \quad (12)$$

The number of cycles required to load scratchpad memories from IFMB, for all the IFM tiles is given by:

$$T_{SP}(i,l) = \Omega(i,l)(d_H(i,l)\,d_V(i,l) + r_{sa}(i) - 1)K \quad (13)$$

where $K$ is used to differentiate between fully connected layers and convolutional layers (i.e. $K = 1$ for fully connected layers and $K = r_f(l)$ for convolutional layers). The time to perform the systolic array computations over all the IFM tiles is given by

$$T_{SA}(i,l) = \Omega(i,l)c_{sa}(i) + T_{SP}(i,l) \quad (14)$$

The time required to write back the OFM for all the tiles back to DRAM is given by

$$T_{out}(i,l) = \frac{1}{W}\alpha(i,l)\,\beta(i,l)(d_H(i,l)d_V(i,l))/s^2(l) \quad (15)$$

So, the total cycles for processing $l^{th}$ layer are and $i^{th}$ design configuration sums up to

$$T(i,l) = T_{FM}(i,l) + T_w(i,l) + T_{sp}(i,l) + T_{sa}(i,l) + T_{out}(i,l) \quad (16)$$

The summation of these $T(i,l)$ for all the layers provides the cumulative clock cycles, $T(i)$, required for a design point. The design point with the lowest $T(i)$ shall represent the most suitable configuration for each of the design parameters.

## III. RESULTS AND DISCUSSION

We use our formulated method to map Tiny YOLO object detection [13] onto an Artix7 FPGA with 86K logic slices, 220 DSP units, and 4.9 Mb of block RAM. We utilized equation (1)-(16), to perform design space exploration of 96 design points, specified by F=4, P=6, Q=4 and R=4. This corresponds to multiple design points corresponding to various combinations of $r_t = \{104, 52, 26, 13, 7, 4\}$, $c_{sa} = \{2, 4, 8, 16\}$, and $ch_{sa} = \{2, 4, 8, 16\}$. Figure 3 provides the layer wise memory requirement, resource analysis, performance estimation and the valid design space for feature map reuse (a-d) and filter reuse (e-h) data traversal order. Since Systimator's resource estimation model comprises layer wise computations of the required memory resources, it allows us to analyze the individual memory load of a particular CNN layer on the architecture. In Fig 3 (a and e) we plot the layer wise memory requirement of the best design point for each of the two modes. Fig 3 (d and g) provide the complete design space covered by the 96 design points being explored. For ease of visualization the design points are plotted on the resource parameters (memory and DSP units) in Fig. 3 (b and f). The dashed lines represent the cut-off points for the memory and DSP ($N_{dsp} = 220$) resources. The design points that are above the memory cut-off line and on the left of the DSP cut-off line contribute to the sub-space that can be implemented on the device under consideration (eq. 10). The various colors relate with different choices of $r_t$. In the figure Black and Red color corresponds to the maximum ($r_t = 104$) and minimum ($r_t = 4$) tile size. In Fig. 3 (c and g) we provide the relative performance ranking of each of these selected design points by plotting $T(i)$ against the utilized DSP units. It can be observed that, in general, the design points that correspond to feature map reuse ($\rho = 0$) require higher memory resources, as they have more points below the memory cut-off line. However, in terms of performance the best design point provided by feature map reuse provides a timing of (12.468 Mil. cycles) as compared to that (12.361 Mil. cycles) of filter reuse. From fig(c) it can be attributed to the lower communication overhead in feature map reuse mode in general, since an IFM tile is required to be brought to the on-chip memory only once. We can infer that for Tiny-Yolo network and Artix7 FPGA we can achieve best performance using any type of data reuse strategy with columns of systolic array to be sixteen whereas rows of systolic array to be six.

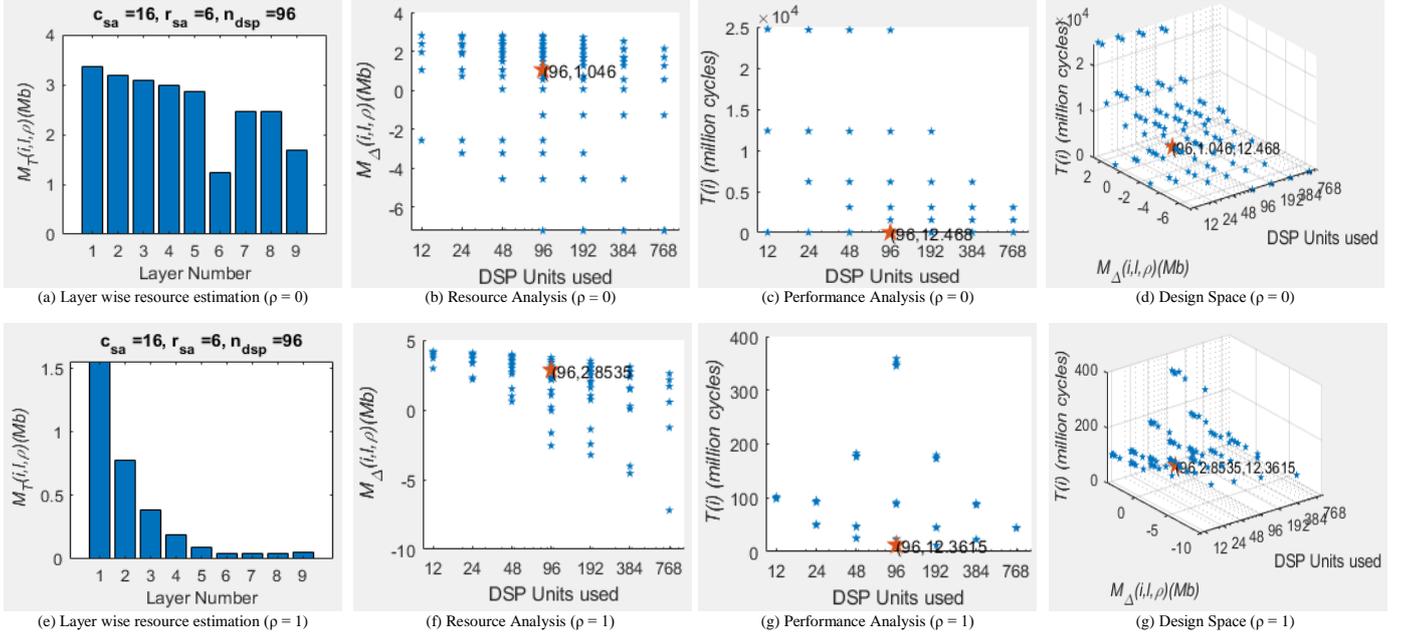

Figure 3: Layer wise memory requirement of a selected design point, resource analysis, performance estimation and valid design space for Feature Map reuse and Filter reuse data traversal orders.

## IV. CONCLUSION AND FUTURE WORK

In this paper we proposed SYStimator, a design space exploration methodology that utilizes analytical models for resource estimation and performance analysis. The methodology relates to a wider class of systolic array based accelerators that take benefit of feature map or filter based design reuse. Our case study on evaluation of convolutional layers of Tiny YOLO demonstrate that the Systimator provides layer-wise in-depth analysis of the required resources and provides an insightful design space exploration. We intend to further evaluate our results with the synthesized results in the future. Furthermore, we shall be improving upon our performance evaluation methodology to incorporate parallel data transfer for IFM and weights.

REFERENCES

[1] Krizhevsky, Alex, Ilya Sutskever, and Geoffrey E. Hinton. "Imagenet classification with deep convolutional neural networks." In Advances in neural information processing systems, pp. 1097-1105. 2012.

[2] Gomes, Tiago, Sandro Pinto, Adriano Tavares, and Jorge Cabral. "Towards an FPGA-based edge device for the Internet of Things." In Emerging Technologies & Factory Automation (ETFA), 2015 IEEE 20th Conference on, pp. 1-4. IEEE, 2015.

[3] Peemen, Maurice, Arnaud AA Setio, Bart Mesman, and Henk Corporaal. "Memory-centric accelerator design for Convolutional Neural Networks." In ICCD, vol. 2013, pp. 13-19. 2013.

[4] Cadambi, Srihari, Abhinandan Majumdar, Michela Becchi, Srimat Chakradhar, and Hans Peter Graf. "A programmable parallel accelerator for learning and classification." In Proceedings of the 19th international conference on Parallel architectures and compilation techniques, pp. 273-284. ACM, 2010.

[5] Farabet, Clément, Cyril Poulet, Jefferson Y. Han, and Yann LeCun. "Cnp: An fpga-based processor for convolutional networks." In Field Programmable Logic and Applications, 2009. FPL 2009. International Conference on, pp. 32-37. IEEE, 2009.

[6] Sankaradas, Murugan, Venkata Jakkula, Srihari Cadambi, Srimat Chakradhar, Igor Durdanovic, Eric Cosatto, and Hans Peter Graf. "A massively parallel coprocessor for convolutional neural networks." In Application-specific Systems, Architectures and Processors, 20th IEEE International Conference on, pp. 53-60. IEEE, 2009.

[7] Chakradhar, Srimat, Murugan Sankaradas, Venkata Jakkula, and Srihari Cadambi. "A dynamically configurable coprocessor for convolutional neural networks." ACM SIGARCH Computer Architecture News 38, no. 3 (2010): 247-257.

[8] Zhang, Chen, Peng Li, Guangyu Sun, Yijin Guan, Bingjun Xiao, and Jason Cong. "Optimizing fpga-based accelerator design for deep convolutional neural networks." In Proceedings of the 2015 ACM/SIGDA International Symposium on Field-Programmable Gate Arrays, pp. 161-170. ACM, 2015.

[9] Wang, Jie, and Jason Cong. "Customizable and high performance matrix multiplication kernel on FPGA." In Proceedings of the ACM/SIGDA International Symposium on Field-Programmable Gate Arrays, pp. 276-276. ACM, 2015.

[10] Zhang, Chen, Zhenman Fang, Peipei Zhou, Peichen Pan, and Jason Cong. "Caffeine: towards uniformed representation and acceleration for deep convolutional neural networks." In Proceedings of the 35th International Conference on Computer-Aided Design, p. 12. ACM, 2016.

[11] Aydonat, Utku, Shane O'Connell, Davor Capalija, Andrew C. Ling, and Gordon R. Chiu. "An OpenCL™ deep learning accelerator on arria 10." In Proceedings of the 2017 ACM/SIGDA International Symposium on Field-Programmable Gate Arrays, pp. 55-64. ACM, 2017.

[12] Wei, Xuechao, Cody Hao Yu, Peng Zhang, Youxiang Chen, Yuxin Wang, Han Hu, Yun Liang, and Jason Cong. "Automated systolic array architecture synthesis for high throughput CNN inference on FPGAs." In Proceedings of the 54th Annual Design Automation Conference 2017, p. 29. ACM, 2017.

[13] Redmon, Joseph, Santosh Divvala, Ross Girshick, and Ali Farhadi. "You only look once: Unified, real-time object detection." In Proceedings of the IEEE conference on computer vision and pattern recognition, pp. 779-788. 2016.

[14] Hazoor Ahmad, 'A Simple Dataflow for Systimator on Tiny-Yolo, AlexNet and VGG16.', Github Inc. URL[https://github.com/hazooree/Systimator-DataFlow].